\documentclass[conference,a4paper]{IEEEtran}
\usepackage{epsfig}
\usepackage{subfigure}
\usepackage{theorem}

\newtheorem{theorem}{Theorem}

\begin{document}

\sloppy

\title{Rate-Distortion Bounds for an $\varepsilon$-Insensitive Distortion Measure} 

\author{
   \IEEEauthorblockN{Kazuho Watanabe}
   \IEEEauthorblockA{Graduate School of Information Science\\
     Nara Institute of Science and Technology,
     Nara, 630-0192, Japan\\
     Email: wkazuho@is.naist.jp} 
 }
\maketitle

\begin{abstract}
Direct evaluation of the rate-distortion function has rarely been achieved when it is strictly greater than its Shannon lower bound. In this paper, we consider the rate-distortion function for the distortion measure defined by an $\varepsilon$-insensitive loss function. We first present the Shannon lower bound applicable to any source distribution with finite differential entropy. Then, focusing on the Laplacian and Gaussian sources, we prove that the rate-distortion functions of these sources are strictly greater than their Shannon lower bounds and obtain analytically evaluable upper bounds for the rate-distortion functions. Small distortion limit and numerical evaluation of the bounds suggest that the Shannon lower bound provides a good approximation to the rate-distortion function for the $\varepsilon$-insensitive distortion measure.
\end{abstract}

\section{Introduction}
In source coding, the rate-distortion function $R(D)$ of a source shows the minimum information rate required to reconstruct the source outputs with average distortion not exceeding $D$.
Rate-distortion functions have been explicitly evaluated for various sources and distortion measures.
The Shannon lower bound (SLB) $R_{L}(D)$ plays an important role in the explicit evaluation of rate-distortion functions of difference distortion measures. A common approach is to derive $R_{L}(D)$ and examine the condition for $R(D)$ to coincide with $R_{L}(D)$ \cite{Cover, Berger}. 
There have been, however, only several results when $R(D)>R_{L}(D)$ for all $D$.
In this case, direct explicit evaluation of $R(D)$ has been achieved only in limited cases such as discrete memoryless finite-alphabet sources \cite{Berger} and a class of sources under an absolute-magnitude distortion measure \cite{TanYao, YaoTan, RD-Gamma}.  
There have also been indirect approaches. Rose proposed a deterministic annealing algorithm to generate $R(D)$ based on the fact that under the squared distortion measure, the optimal reconstruction is purely discrete when $R(D) > R_{L}(D)$.
Buzo {\it{et al.}} obtained upper and lower bounds for $R(D)$ under the Itakura-Saito distortion measure \cite{Buzo}.

In this paper, we focus on the $\varepsilon$-insensitive loss function as a distortion measure, which was introduced to support vector machines for regression \cite{Vapnik}. We obtain the SLB for this difference distortion measure, which is analytically evaluable for arbitrary sources with finite differential entropy.
Then, we examine the condition for the rate-distortion function to coincide with the SLB.
Taking the Laplacian and Gaussian sources as specific examples, we prove that the rate-distortion functions of these sources lie strictly above their SLBs for all $D$ when $\varepsilon >0$ and derive analytically evaluable upper bounds for the rate-distortion functions.
Investigation of small distortion limit of these upper bounds shows that the SLB has the accuracy of $O(\varepsilon^{2})$ as $D\rightarrow 0$ in both sources.
Numerical evaluation of the lower and upper bounds demonstrates that the SLB gives a good approximation to $R(D)$ for small distortion while the trivial upper bound provided by the rate-distortion function for $\varepsilon=0$ suggests that the SLB also gives a reasonable approximation for large distortion as well. 

\section{Rate-Distortion Function for the $\varepsilon$-Insensitive Distortion Measure}
\subsection{Rate-Distortion Function}
Let $X$ and $Y$ be random variables on $\mathbf{R}$ and $d(x,y)$ be the non-negative distortion measure between $x$ and $y$.
The rate-distortion function $R(D)$ of the source $X\sim p(x)$ with respect to the distortion $d$ is defined by
\begin{equation}
R(D) = \inf_{q(y|x) : E[d(x,y)] \leq D } I(q), \label{eq:rate_dist}
\end{equation}
where
\begin{eqnarray*}
I(q) 
&=&\int \int q(y|x)p(x)\log\frac{q(y|x)}{\int q(y|x) p(x) dx} dx dy 
\end{eqnarray*}
is the mutual information and $E$ denotes the expectation with respect to $q(y|x)p(x)$.
$R(D)$ shows the minimum achievable rate for the i.i.d. source with the density $p(x)$ under the given distortion measure $d$ \cite{Berger, Cover}.

The above minimization problem can be reformulated as a problem of minimization over the reproduction density $q(y)$ \cite{Berger, Gray},
\begin{equation}
 \inf_{q(y)}\left[-\int p(x)\log \int\exp(sd(x,y))q(y)dydx\right], \label{eq:rd_recon}
\end{equation}
where $s\leq 0$ is a parameter.
Then, if there exists $q_{s}(y)$ that achieves the infimum in Eq.~(\ref{eq:rd_recon}), $R(D)$ is parametrically given by
\begin{eqnarray}
R(D_{s}) \!\!\!\!& = & \!\!\! -\int p(x)\log \int\exp(sd(x,y))q_{s}(y)dydx+ sD_{s}, \nonumber\\
D_{s} &=& \int\int p(x)q_{s}(y|x)d(x,y)dxdy, \label{eq:rd_para}
\end{eqnarray}
where the optimal conditional density of reconstruction, $q_{s}(y|x)$ is defined by
\begin{equation}
q_s(y|x)=\frac{ q_{s}(y)\exp(sd(x,y)) }{ \int q_{s}(y)\exp(sd(x,y)) dy}. \label{eq:opt_cond}
\end{equation}
In Eq.~(\ref{eq:rd_para}), $R(D)$ is parameterized by $s\leq 0$, which corresponds to the slope of the tangent of $R(D)$ at $(D_{s}, R(D_{s}))$ and hence is referred to as the slope parameter \cite{Berger}.

From the properties of the rate-distortion function $R(D)$, we know that $R(D)>0$ for $0<D<D_{\max}$, where 
\begin{equation}
D_{\max} = \inf_{y} \int p(x) d(x, y) dx, \label{eq:D_max}
\end{equation}
and $R(D)=0$ for $D\geq D_{\max}$ \cite[p.~90]{Berger}. 

\subsection{$\varepsilon$-Insensitive Loss Function}
\begin{figure}[ht]
\begin{center}
\hspace{-4.5mm}
\vspace{-5mm}
\includegraphics[width=0.35\textwidth]{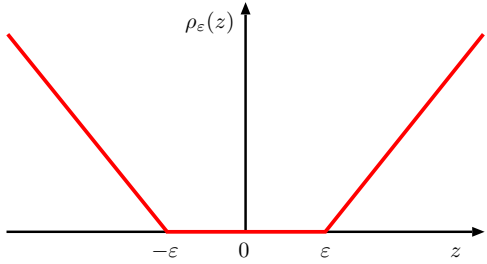} 
\end{center}
\caption{The $\varepsilon$-insensitive loss function $\rho_{\varepsilon}(z)$.}
\label{fig:epsi-loss}
\end{figure}

In this paper, we focus on the following difference distortion measure defined by the $\varepsilon$-insensitive loss function $\rho_{\varepsilon}$ (Fig. \ref{fig:epsi-loss}),
\begin{equation}
d(x, y) = \rho_{\varepsilon} (x-y), \label{eq:dist_epsi}
\end{equation}
where
\[
\rho_{\varepsilon}(z) = \left\{
\begin{array}{ll}
 |z|-\varepsilon, &(|z| \geq \varepsilon),
\\ 
 0, &(|z| < \varepsilon).
\end{array}
\right.  
\]
\vspace{-5mm}

This loss function with $\varepsilon > 0$ was introduced to support vector regression in order to provide a sparsity inducing mechanism \cite{Vapnik,Steinw}.
We denote the rate-distortion function for this distortion measure by $R^{(\varepsilon)}(D)$ and the maximum distortion $D_{\max}$ in Eq.~(\ref{eq:D_max}) by $D_{\max}^{(\varepsilon)}$.

\subsection{Shannon Lower Bound}
\label{sec:SLB}
Generally for difference distortion measures, Shannon obtained a lower bound to $R(D)$, which is referred to as the Shannon lower bound (SLB) \cite[p.~92]{Berger}.
For the $\varepsilon$-insensitive distortion measure, it is parametrically expressed as
\begin{eqnarray}
R^{(\varepsilon)}(D_{s}) \geq  R_{L}^{(\varepsilon)}(D_{s}) &=& h(p) -h(g_{s}), \label{eq:SLB} \\
D_{s} &=& \int \rho_{\varepsilon}(x)g_{s}(x)dx, \label{eq:SLB_dist}
\end{eqnarray}
where $h(p)=-\int p(x)\log p(x)dx$ is the differential entropy of the probability density $p$ and $g_{s}$ is the probability density function defined by\footnote{
We omit the dependency on $\varepsilon$ in notations unless we put $\varepsilon=0$.
}
\begin{equation}
g_{s}(x) = \frac{ e^{s\rho_{\varepsilon}(x)}}{\int e^{s\rho_{\varepsilon}(z)} dz}. \label{eq:kern}
\end{equation}

We explicitly evaluate $h(g_{s})$ to obtain the SLB.
The density $g_{s}$ is explicitly given by
\begin{equation}
g_{s}(x) = \left\{
\begin{array}{ll}
 \frac{1}{C_{s}}e^{-s(x+\varepsilon)}, &(x \leq -\varepsilon),
\\ 
 \frac{1}{C_{s}}, &(|x| < \varepsilon),
\\
 \frac{1}{C_{s}}e^{s(x-\varepsilon)}, &(x \geq \varepsilon),
\end{array}
\right.   \label{eq:kern_epsi}
\end{equation}
where
\begin{equation}
C_{s} = 2\frac{1+|s|\varepsilon}{|s|}. \label{eq:const_epsi}
\end{equation}
Its differential entropy is evaluated as,
\begin{eqnarray}
h(g_{s}) 
&=& \log\left(2\frac{1+|s|\varepsilon}{|s|}\right) + \frac{1}{1+|s|\varepsilon}. \label{eq:ent_gs}
\end{eqnarray}
The slope parameter $s$ is related to the average distortion $D_{s}$ by Eq.~(\ref{eq:SLB_dist}), which is rewritten as,
\begin{equation}
D_{s} = \frac{2}{s^{2}C_{s}} =\frac{1}{(1+\varepsilon |s|)|s|}. \label{eq:SLB_dist_epsi}
\end{equation}
Solving this for $|s|$ yields
\[
|s| = \frac{-D_{s}+\sqrt{D_{s}^{2}+4D_{s}\varepsilon}}{2D_{s}\varepsilon}.
\]
Putting this back into Eq.~(\ref{eq:ent_gs}), we have
\[
h(g_{s}) 
= \log(2\varepsilon +D_{s}+ \sqrt{D_{s}^{2}+4D_{s}\varepsilon})
-\frac{D_{s}- \sqrt{D_{s}^{2}+4D_{s}\varepsilon}}{2\varepsilon}.
\]
Thus, from Eq.~(\ref{eq:SLB}), we obtain the following theorem.
\begin{theorem}
\label{thm:SLB_epsi}
The rate-distortion function for the $\varepsilon$-insensitive distortion measure in Eq.~(\ref{eq:dist_epsi}) satisfies $R^{(\varepsilon)}(D) \geq R_{L}^{(\varepsilon)}(D)$ for all $D$, where 
\begin{eqnarray*}
R_{L}^{(\varepsilon)}(D) &=& h(p) -\log(2\varepsilon) -\log\left(1+\tilde{D}+ \sqrt{\tilde{D}^{2}+2\tilde{D}}\right) \\&&+\tilde{D} - \sqrt{\tilde{D}^{2}+2\tilde{D}}, 
\end{eqnarray*}
$\tilde{D}=\frac{D}{2\varepsilon}$ and $h(p)$ is the differential entropy of the source density.
\end{theorem}

\subsection{Condition for $R^{(\varepsilon)}(D) = R_{L}^{(\varepsilon)}(D)$}
For any negative value of the slope parameter $s$, the lower bound $R_{L}^{(\varepsilon)}(D_{s})$ coincides with $R^{(\varepsilon)}(D_{s})$ if and only if the condition
\begin{equation}
p(x) = \int q(y) g_{s}(x-y)dy, \label{eq:cond_SLB}
\end{equation}
holds for all $x$ and a valid density function $q(y)$ \cite[p.~94]{Berger}.
The condition in Eq.~(\ref{eq:cond_SLB}) is equivalent to 
\begin{equation}
P(\omega) = Q(\omega)G_{s}(\omega), \label{eq:cond_SLB_F}
\end{equation} 
where $P$, $Q$ and $G_{s}$ are the Fourier transforms (characteristic functions) of $p$, $q$ and $g_{s}$ respectively.

The Fourier transform of $g_{s}$ in Eq.~(\ref{eq:kern_epsi}) is specifically given by
\begin{eqnarray}
G_{s}(\omega) &=& \int \frac{e^{s\rho_{\varepsilon}(x)}}{C_{s}}e^{-i\omega x}dx \nonumber\\
&=& \frac{s^{2}}{s^{2}+\omega^{2}} \cdot \frac{\varepsilon |s| \frac{\sin(\omega \varepsilon)}{\omega \varepsilon} +\cos(\omega\varepsilon)}{1+|s|\varepsilon} \nonumber\\
&\equiv & L_{|s|}(\omega)\cdot M_{|s|}^{(\varepsilon)}(\omega). \label{eq:kern_epsi_F}
\end{eqnarray}
Here, the first factor, defined as $L_{|s|}(\omega) = \frac{s^{2}}{s^{2}+\omega^{2}}$, is the characteristic function of the Laplace distribution with parameter $|s|$ whose density function is $l_{|s|}(x)=\frac{|s|}{2}e^{s|x|}$. The second factor, $M_{|s|}^{(\varepsilon)}(\omega) = \frac{\varepsilon |s| \frac{\sin(\omega \varepsilon)}{\omega \varepsilon} +\cos(\omega\varepsilon)}{1+|s|\varepsilon} $, is the characteristic function of the mixture of the delta distributions (on $-\varepsilon$ and $\varepsilon$) and the uniform distribution on $[-\varepsilon, \varepsilon]$ mixed with the proportion $1:\varepsilon |s|$. 
More specifically, the density function of this mixture is expressed as
\begin{eqnarray*}
m_{|s|}^{(\varepsilon)}(x) 
= \frac{1}{1+\varepsilon |s|} \frac{\delta(x-\varepsilon)+\delta(x+\varepsilon)}{2} 
\! + \!\frac{\varepsilon |s|}{1+\varepsilon |s|} u_{[-\varepsilon, \varepsilon]}(x), 
\end{eqnarray*}
where $\delta$ is the Dirac delta function and $u_{[-\varepsilon, \varepsilon]}$ is the density function of the uniform distribution on $[-\varepsilon, \varepsilon]$.
Hence, Eq.~(\ref{eq:kern_epsi_F}) means that the density $g_{s}$ is given by the convolution $l_{|s|}*m_{|s|}^{(\varepsilon)}$ of $l_{|s|}$ and $m_{|s|}^{(\varepsilon)}$. 
Summarizing Eqs.~(\ref{eq:cond_SLB_F}) and (\ref{eq:kern_epsi_F}), we see that for the $\varepsilon$-insensitive distortion measure, the condition for $R_{L}^{(\varepsilon)}(D)$ to coincide with $R^{(\varepsilon)}(D)$ is the existence of a valid characteristic function $Q(\omega)$ satisfying 
\begin{equation}
P(\omega)=Q(\omega)L_{|s|}(\omega)M_{|s|}^{(\varepsilon)}(\omega) \label{eq:cond_SLB_F_summ},
\end{equation}
for the characteristic function $P(\omega)$ of the source distribution.

The above condition is rewritten as $P(\omega)/L_{|s|}(\omega)=M_{|s|}^{(\varepsilon)}(\omega)Q(\omega)$.
If $Q(\omega)$ is the characteristic function of a probability distribution, that is, the Fourier transform of a density $q(y)$, then the right hand side is that of convolution of $q$ and $m_{|s|}^{(\varepsilon)}$.
Hence, Eq.~(\ref{eq:cond_SLB_F_summ}) is never satisfied for any density $q(y)$ unless $P(\omega)/L_{|s|}(\omega)$ is the Fourier transform of a valid density.
This means that  a necessary condition for $R^{(\varepsilon)}(D_{s})=R_{L}^{(\varepsilon)}(D_{s})$ is given by $R^{(0)}(D_{s})=R_{L}^{(0)}(D_{s})$, that is, the SLB coincides with the rate-distortion function under the absolute-magnitude distortion ($\varepsilon =0$). 

We will use the above condition in Section \ref{sec:LapGau} to prove $R_{L}^{(\varepsilon)}(D)$ is strictly smaller than $R^{(\varepsilon)}(D)$ for the Laplacian and Gaussian sources.

\subsection{General Upper Bound}
Let us turn to upper bounds for $R^{(\varepsilon)}(D)$.
Since $\rho_{\varepsilon}(x)\leq \rho_{0}(x) = |x|$, we have a trivial upper bound,
\[
R^{(\varepsilon)}(D) \leq R^{(0)}(D),
\]
where $R^{(0)}(D)$ is the rate-distortion function for the absolute-magnitude distortion measure, $d(x,y)=|x-y|$.

Another more informative upper bound is obtained by taking $q(y|x)=g_{s}(y-x)$, where $g_{s}$ is defined by Eq.~(\ref{eq:kern}), in the original rate-distortion problem in Eq.~(\ref{eq:rate_dist}) \cite[p.~103]{Berger}.
This yields the following upper bound,
\begin{equation}
R^{(\varepsilon)}(D_{s}) \leq R_{U}^{(\varepsilon)}(D_{s}) = h(r_{s})-h(g_{s}), \label{eq:gene_up}
\end{equation}
where 
\begin{equation}
r_{s}(y) = (g_{s}*p)(y) = \int g_{s}(y-x)p(x)dx \label{eq:marg_recon}
\end{equation}
and $D_{s}$ is given by Eq.~(\ref{eq:SLB_dist}) and further by Eq.~(\ref{eq:SLB_dist_epsi}).
Note in Eq.~(\ref{eq:gene_up}) that the term $h(g_{s})$ is common to the SLB and is specifically given by Eq.~(\ref{eq:ent_gs}). 

Since $g_{s}$ is defined by $\rho_{\varepsilon}$ as in Eq.~(\ref{eq:kern}), $h(r_{s})$ is analytically intractable for many sources. Hence, we create a further upper bound which is analytically obtained for any souces with finite variance.

The Gaussian distribution with variance $v$ maximizes differential entropy among the distributions whose variance is constrained to be $v$.
The maximum value of differential entropy is $\frac{1}{2}\log(2\pi e v)$.
Therefore, the differential entropy $h(r_{s})=h(g_{s}*p)$ is upper bounded as follows,
\begin{equation}
h(r_{s})\leq \frac{1}{2}\log\left(2\pi e(v_{p}+v_{s}^{(\varepsilon)})\right), \label{eq:gau_bou}
\end{equation}  
where $v_{p}= \int x^{2}p(x)dx - \left(\int x p(x) dx\right)^{2}$ and $v_{s}^{(\varepsilon)}= \int x^{2}g_{s}(x)dx$.
This is because the variance of $g_{s}*p$ is $v_{p}+v_{s}^{(\varepsilon)}$. The variance of $g_{s}$ is specifically evaluated as,
\begin{equation}
v_{s}^{(\varepsilon)} = \frac{2}{C_{s}}\left\{ \frac{\varepsilon^{3}}{3}+\frac{1}{|s|} \left(\varepsilon^{2}+\frac{2}{|s|}\varepsilon+\frac{2}{|s|^{2}}\right)\right\}, \label{eq:var_gs}
\end{equation}
where $C_{s}$ is defined in Eq.~(\ref{eq:const_epsi}).
The general upper bound in Eq.~(\ref{eq:gene_up}), combined with  Eqs.~(\ref{eq:gau_bou}) and (\ref{eq:var_gs}), yields the following upper bound to $R_{U}^{(\varepsilon)}(D)$,
\begin{equation}
R_{GE}^{(\varepsilon)}(D_{s})= \frac{1}{2}\log\left(2\pi  e(v_{p}+v_{s}^{(\varepsilon)})\right) -h(g_{s}), \label{eq:gauent_up}
\end{equation}
which is referred to as the Gaussian entropy bound.

In the next sections, we will evaluate these upper bounds for the Laplacian and Gaussian sources to examine the tightness of the general lower bound obtained in Theorem \ref{thm:SLB_epsi}.

\section{Laplacian and Gaussian Sources}
\label{sec:LapGau}
\subsection{Laplacian Source}
In this subsection, we consider the Laplacian source with parameter $\alpha$,
\begin{equation}
p(x) = l_{\alpha}(x) = \frac{\alpha}{2}e^{-\alpha |x|}. \label{eq:Laplace}
\end{equation}
The SLB for this source is given by Theorem \ref{thm:SLB_epsi} with the differential entropy,
\[
h(p)= 1-\log\frac{\alpha}{2}.
\]
The maximum distortion in Eq.~(\ref{eq:D_max}) is 
\begin{equation}
D_{\max}^{(\varepsilon)}=\int \rho_{\varepsilon}(x)p(x)dx=\frac{1}{\alpha}e^{-\alpha\varepsilon}. \label{eq:D_max_lap}
\end{equation}
For the absolute-magnitude distortion measure ($\varepsilon=0$), 
\begin{equation}
R^{(0)}(D)=R_{L}^{(0)}(D)= -\log(\alpha D), \;\;(0 \leq D \leq 1/\alpha),
   \label{eq:RD_abs_lap}
\end{equation}
holds \cite[p.~95, Example 4.3.2.1]{Berger} because the condition in Eq.~(\ref{eq:cond_SLB_F_summ}) reduces to $M_{|s|}^{(0)}(\omega)=1$ and 
\[
Q(\omega)=\frac{\alpha^{2}}{|s|^{2}} + \left(1 -\frac{\alpha^{2}}{|s|^{2}}  \right)\frac{\alpha^{2}}{\alpha^{2}+\omega^{2}},
\]
which is the Fourier transform of the valid probability density,
$q(y)=\frac{\alpha^{2}}{|s|^{2}}\delta(y) +  \left(1 -\frac{\alpha^{2}}{|s|^{2}}  \right)l_{\alpha}(y)$.
For $\varepsilon > 0$, however, $R^{(\varepsilon)}(D)$ is strictly greater than $R_{L}^{(\varepsilon)}(D)$ for all $D$, which we will prove in the following.  We will later derive an analytic upper bound to $R^{(\varepsilon)}(D)$ from Eq.~(\ref{eq:gene_up}).

The condition for $R^{(\varepsilon)}(D)=R_{L}^{(\varepsilon)}(D)$ in Eq.~(\ref{eq:cond_SLB_F_summ}) is equivalent to
\[
Q(\omega)=\frac{1}{M_{|s|}^{(\varepsilon)}(\omega)} \left\{ \frac{\alpha^{2}}{|s|^{2}} +\left(1- \frac{\alpha^{2}}{|s|^{2}} \right)\frac{\alpha^{2}}{\alpha^{2}+\omega^{2}} \right\}. 
\]
For $|s|>\alpha$, $\left\{ \frac{\alpha^{2}}{|s|^{2}} +\left(1- \frac{\alpha^{2}}{|s|^{2}} \right)\frac{\alpha^{2}}{\alpha^{2}+\omega^{2}} \right\} > \frac{\alpha^{2}}{|s|^{2}}$. Putting $\omega=(2k-1/2)\frac{\pi}{\varepsilon}$, where $k$ is a natural number, we have
\[
\left|Q\left((2k-1/2)\frac{\pi}{\varepsilon}\right)\right| > \frac{\alpha^{2}}{|s|^{2}}\frac{1+\varepsilon |s|}{\varepsilon |s|}  (2k-1/2)\pi . 
\]
Since the right hand side becomes arbitrarily large for $k\rightarrow \infty$,
$Q(\omega)$ can not be a Fourier transform of any density $q(y)$. 
This means that $R_{L}^{(\varepsilon)}(D_{s})<R^{(\varepsilon)}(D_{s})$ for $s<-\alpha$. 
It follows from Eq.~(\ref{eq:ent_gs}) that
$R_{L}^{(\varepsilon)}(D_{s=-\alpha}) = 1-\log(1+\alpha\varepsilon)-\frac{1}{1+\alpha\varepsilon} \leq 0$, which implies that $R^{(\varepsilon)}(D)$ is strictly greater than $R_{L}^{(\varepsilon)}(D)$ for all $D$.

To obtain an analytic upper bound for $R^{(\varepsilon)}(D)$, we evaluate $h(r_{s})$ in Eq.~(\ref{eq:gene_up}). The density $r_{s}(y)$ defined in Eq.~(\ref{eq:marg_recon}) is specifically given by
\[
r_{s}(y) =  \left\{
\begin{array}{ll}
 \frac{1}{2C_{s}}b_{s}(-y), &(y \leq -\varepsilon),
\\ 
 \frac{1}{2C_{s}}a_{s}(y), &(|y| < \varepsilon),
\\
 \frac{1}{2C_{s}}b_{s}(y), &(y \geq \varepsilon),
\end{array}
\right.   
\] 
 where 
$a_{s}(y) = \frac{s}{\alpha -s}e^{-\alpha (y+\varepsilon)}+\frac{s}{\alpha -s}e^{\alpha (y-\varepsilon)}+2$ and $b_{s}(y) = \frac{s}{\alpha -s}e^{-\alpha (y+\varepsilon)}+\frac{s}{\alpha + s}e^{-\alpha (y-\varepsilon)}+\frac{2\alpha^{2}}{\alpha^{2}-s^{2}}e^{s(y-\varepsilon)}$.
Since the differential entropy of $r_{s}(y)$ is not analytically simplified any more, we evaluate it from above to obtain an upper bound for $R^{(\varepsilon)}(D_{s})$.

First, we bound $a_{s}(y)$ and $b_{s}(y)$ from below.
\[
a_{s}(y) \geq 2+ \frac{2se^{-\alpha\varepsilon}}{\alpha-s}\cosh(\alpha\varepsilon) \equiv c_{s}
\]
for $|y|<\varepsilon$ and since $s>-\alpha$, 
\begin{eqnarray*}
b_{s}(y) \!\!\!\!\! &\geq & \!\!\!\!\! \frac{s}{\alpha -s}e^{-\alpha (y+\varepsilon)}+\frac{s}{\alpha + s}e^{-\alpha (y-\varepsilon)}+\frac{2\alpha^{2}}{\alpha^{2}-s^{2}}e^{-\alpha(y-\varepsilon)} \\
&=& c_{s}e^{\alpha (\varepsilon-y)} 
\end{eqnarray*}
for $y \geq \varepsilon$.
Next, let us define 
\[
B_{s} \equiv \int_{\varepsilon}^{\infty} b_{s}(y) dy 
= \frac{s}{\alpha -s}\frac{1}{\alpha} e^{-2\alpha\varepsilon} +\frac{s^{2}(\alpha -s)-2\alpha^{3}}{(\alpha^{2}-s^{2})s\alpha}.
\]
Finally, $h(r_{s})$ is bounded as follows,
\begin{equation}
h(r_{s}) \leq -\log\frac{c_{s}}{2C_{s}} -\frac{\alpha\varepsilon}{C_{s}}B_{s} + \frac{\alpha}{C_{s}}E_{s}, \label{eq:ent_rs_up}
\end{equation}
where 
\begin{eqnarray*}
&&E_{s}\equiv\int_{\varepsilon}^{\infty} y b_{s}(y) dy\\
&=& \frac{s}{\alpha -s}\frac{1+\alpha\varepsilon}{\alpha^{2}}e^{-2\alpha\varepsilon}+\frac{s}{s+\alpha}\frac{1+\alpha\varepsilon}{\alpha^{2}}+\frac{2\alpha^{2}}{\alpha^{2}-s^{2}}
\frac{1-s\varepsilon}{s^{2}},
\end{eqnarray*}
which easily follows from $\int_{\varepsilon}^{\infty}ye^{-\alpha y}dy = \frac{1+\alpha\varepsilon}{\alpha^{2}}e^{-\alpha\varepsilon}$.

Thus, we have the further upper bound, $R_{U}^{(\varepsilon)}(D_{s}) \leq R_{AU}^{(\varepsilon)}(D_{s})$, where
\begin{equation}
R_{AU}^{(\varepsilon)}(D_{s}) \equiv -\log\frac{c_{s}}{2C_{s}} -\frac{\alpha\varepsilon}{C_{s}}B_{s} + \frac{\alpha}{C_{s}}E_{s} -h(g_{s}), \label{eq:anal_up}
\end{equation}
which we will refer to as the analytic upper bound in Section \ref{sec:numeri}.

In the low distortion limit, $D\rightarrow 0$ and $|s|\rightarrow \infty$, we have
$C_{s} \rightarrow  2\varepsilon$,  $c_{s} \rightarrow 2e^{-\alpha\varepsilon} \sinh(\alpha\varepsilon)$, 
$B_{s} \rightarrow \frac{2}{\alpha}e^{-\alpha\varepsilon} \sinh(\alpha\varepsilon)$, 
$E_{s} \rightarrow \frac{1+\alpha\varepsilon}{\alpha^{2}}2e^{-\alpha\varepsilon} \sinh(\alpha\varepsilon)$. 
Then, Eq.~(\ref{eq:ent_rs_up}) turns out to be
\[
h(r_{-\infty}) \leq \alpha\varepsilon -\log\frac{\sinh(\alpha\varepsilon)}{2\varepsilon}+\frac{e^{-\alpha\varepsilon}}{\alpha\varepsilon}\sinh(\alpha\varepsilon).
\]
Furthermore, it follows from $\frac{\sinh(\alpha\varepsilon)}{\alpha\varepsilon} = 1+\frac{(\alpha\varepsilon)^{2}}{3!}+O(\varepsilon^{4})$ and $e^{-\alpha\varepsilon} = 1-\alpha\varepsilon+\frac{(\alpha\varepsilon)^{2}}{2}+O(\varepsilon^{3})$ that
\[
h(r_{-\infty}) \leq 
h(p) +\frac{(\alpha\varepsilon)^{2}}{2}+ O(\varepsilon^{3}),
\]
which implies that the SLB $R_{L}^{(\varepsilon)}(D)$ satisfies
$R_{L}^{(\varepsilon)}(0) \leq R^{(\varepsilon)}(0) \leq R_{L}^{(\varepsilon)}(0) + \frac{(\alpha\varepsilon)^{2}}{2}+ O(\varepsilon^{3})$.

\subsection{Gaussian Source}
In this subsection, we consider the Gaussian source with mean zero and variance $\sigma^{2}$,
\begin{equation}
p(x) = \sqrt{\frac{1}{2\pi \sigma^{2}}}e^{-\frac{x^{2}}{2\sigma^{2}}}. \label{eq:Gauss}
\end{equation}
The differential entropy of this source is $h(p)=\frac{1}{2}\{1+\log(2\pi \sigma^{2})\}$. The maximum distortion in Eq.~(\ref{eq:D_max}) is 
\begin{equation}
D_{\max}^{(\varepsilon)}=\int\rho_{\varepsilon} (x) p(x) dx = 2\left\{\sigma^{2} p(\varepsilon) -\varepsilon \Phi_{c}(\varepsilon/\sigma)\right\}, \label{eq:D_max_gau}
\end{equation}
where $\Phi_{c}(x)=\frac{1}{\sqrt{2\pi}}\int_{x}^{\infty}e^{-t^{2}/2}dt$.
We show that $R^{(\varepsilon)}(D)$ for this source lies above its SLB for all $D$ and consider small distortion limit of the Gaussian entropy upper bound in Eq.~(\ref{eq:gauent_up}).

The condition for $R^{(\varepsilon)}(D_{s})=R_{L}^{(\varepsilon)}(D_{s})$ is given by Eq.~(\ref{eq:cond_SLB_F_summ}) with $P(\omega)=e^{-\sigma^{2}\omega^{2}/2}$.
The inverse transform of $P(\omega)/L_{|s|}(\omega)=\left(1+\frac{\omega^{2}}{|s|^{2}}\right)P(\omega)$ is 
$p(x)\left(1+\frac{1}{|s|^{2}\sigma^{2}}-\frac{x^{2}}{|s|^{2}\sigma^{4}}\right)$,
which becomes negative for large $x$.\footnote{See also Eq.~(4.~3.~27) in \cite[p.~97]{Berger}.}
From the discussion below Eq.~(\ref{eq:cond_SLB_F_summ}), this implies that $R^{(0)}(D) > R_{L}^{(0)}(D)$ and hence $R^{(\varepsilon)}(D) > R_{L}^{(\varepsilon)}(D)$ for all $D$.  

We have the Gaussian entropy bound in Eq.~(\ref{eq:gauent_up}) with $v_{p}=\sigma^{2}$.
In the limit, $|s|\rightarrow \infty$, $v_{s}^{(\varepsilon)}\rightarrow \varepsilon^{2}/3$. Then, the upper bound in Eq.~(\ref{eq:gau_bou}) is further upper bounded as follows,
\[
 \frac{1}{2}\log\left(2\pi e\left(\sigma^{2}+\frac{\varepsilon^{2}}{3}\right)\right)
\leq h(p) + \frac{\varepsilon^{2}}{6\sigma^{2}},
\]
since $\log(1+ x) \leq x$ for $x>0$.
This means that the SLB given by Theorem \ref{thm:SLB_epsi} provides an approximation to $R^{(\varepsilon)}(D)$ with accuracy $\frac{\varepsilon^{2}}{6\sigma^{2}}$ as $D\rightarrow 0$.

\section{Numerical Evaluation}
\label{sec:numeri}

\begin{figure}[t!]
\begin{center}
\includegraphics[width=0.5\textwidth]{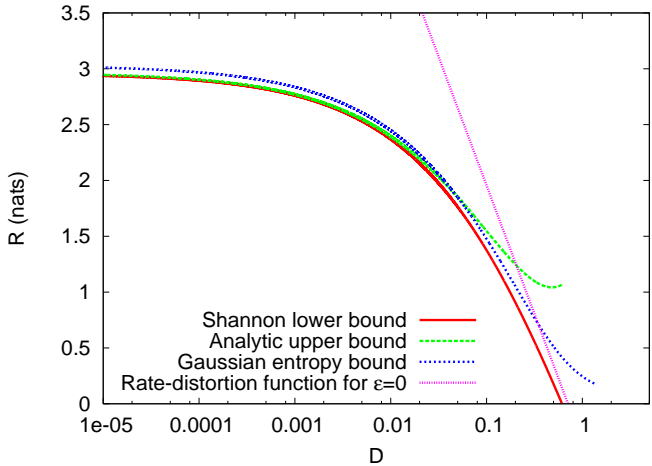} 
\end{center}
\vspace{-5mm}
\caption{Rate-distortion bounds for the Laplacian source. }
\label{fig:rd-epsi-lap}
\end{figure}

Figure \ref{fig:rd-epsi-lap} depicts the functions $R_{L}^{(\varepsilon)}(D)$ and $R_{AU}^{(\varepsilon)}(D)$ for the Laplacian source in Eq.~(\ref{eq:Laplace}) with $\alpha=\sqrt{2}$ when $\varepsilon = 0.1$.
It also shows $R_{GE}^{(\varepsilon)}(D)$ in Eq.~(\ref{eq:gauent_up}) with $v_{p}=2/\alpha^{2}=1$ and the trivial upper bound $R^{(0)}(D)$ given by Eq.~(\ref{eq:RD_abs_lap}).
It is observed that the analytic upper bound and the SLB are very close to each other for small distortion ($D<0.01$). The analytic upper bound becomes looser than the Gaussian entropy bound for large distortion ($0.05<D$) while the trivial upper bound is relatively more informative about $R^{(\varepsilon)}(D)$ in the vicinity of $D_{\max}^{(\varepsilon)}$ when combined with the SLB.
This suggests that the SLB provides reasonable approximation to $R^{(\varepsilon)}(D)$ even for large distortion. Let $D_{\max}^{(\varepsilon), L}$ denote the average distortion where the SLB $R_{L}^{(\varepsilon)}(D)$ reaches zero. We observed the following values of $D_{\max}^{(\varepsilon), L}$,  $D_{\max}^{(\varepsilon)}$ defined in Eq.~(\ref{eq:D_max_lap}) and  $D_{\max}^{(0)}=1/\alpha$,
$D_{\max}^{(\varepsilon), L} = 0.6136 < D_{\max}^{(\varepsilon)}=0.6139 <  D_{\max}^{(0)}=0.7071$.   

\begin{figure}[t]
\begin{center}
\includegraphics[width=0.5\textwidth]{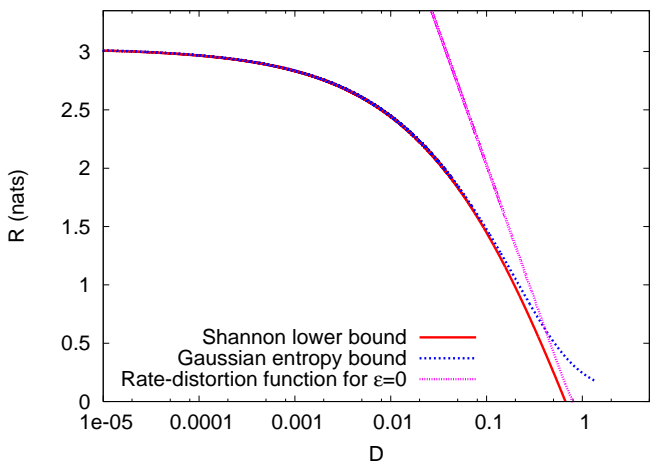} 
\end{center}
\vspace{-5mm}
\caption{Rate-distortion bounds for the Gaussian source.}
\label{fig:rd-epsi-gau}
\end{figure}

Figure \ref{fig:rd-epsi-gau} presents the bounds $R_{L}^{(\varepsilon)}(D)$ and $R_{GE}^{(\varepsilon)}(D)$ ($\varepsilon = 0.1$) and $R^{(0)}(D)$ for the Gaussian source in Eq.~(\ref{eq:Gauss}) with $\sigma^{2}=1$. The rate-distortion function $R^{(0)}(D)$ for the Gaussian source was evaluated according to its explicit parametric form given in \cite[Th. 2]{TanYao}.
It can be seen that the Gaussian entropy bound is very tight for small distortion ($D<0.1$), which implies high accuracy of the SLB and that the trivial upper bound is informative about $R^{(\varepsilon)}(D)$ around $D=D_{\max}^{(\varepsilon)}$.
The observed values of $D_{\max}^{(\varepsilon), L}$,  $D_{\max}^{(\varepsilon)}$ defined in Eq.~(\ref{eq:D_max_gau}) and  $D_{\max}^{(0)}=\sqrt{2\sigma^{2}/\pi}$ are as follows,
$D_{\max}^{(\varepsilon), L} = 0.6662 < D_{\max}^{(\varepsilon)}=0.7019 <  D_{\max}^{(0)}=0.7979$. 

\section{Conclusion}
In this article, we have shown upper and lower bounds for the rate-distortion function of the $\varepsilon$-insensitive distortion measure. 
We derived the SLB, which is applicable to any source densities.
Focusing on the Laplacian and Gaussian sources, we have proved that the rate-distortion functions for these sources are strictly greater than the corresponding SLBs for all $D$ and provided upper bounds for the rate-distortion functions, which are proved to have accuracy of $O(\varepsilon^{2})$ in the small distortion limit.
We have demonstrated through numerical evaluation that the SLB is very accurate in the small distortion region while it still provides reasonable approximation to $R^{(\varepsilon)}(D)$ for the high distortion region around $D_{\max}^{(\varepsilon)}$ as $R^{(0)}(D)$ suggests.

In order to explicitly evaluate $R^{(\varepsilon)}(D)$, it would be important to explore properties of the optimal reproduction distribution $q_{s}(y)$. It is also important to develop practical learning algorithm for the mixture $\int g_{s}(x-y)q(y)dy$, where $g_{s}$ is defined by the $\varepsilon$-insensitive loss function as in Eq.~(\ref{eq:kern_epsi}).
Another issue to be addressed is the extension of our results to vector sources by using an extension of the $\varepsilon$-insensitive loss function to higher-dimensional vectors.
There are some variations of the $\varepsilon$-insensitive loss function \cite{Chu, Dekel}. It would also be an interesting undertaking to investigate properties of the rate-distortion functions for these variations.

\section*{Acknowledgments}
This work was supported by JSPS KAKENHI Grant Number 23700175.


\end{document}